\documentstyle[12pt,epsfig]{article}
\newcommand{\beq}{\begin{equation}}
\newcommand{\eeq}{\end{equation}}
\newcommand{\beqs}{\begin{eqnarray}}
\newcommand{\eeqs}{\end{eqnarray}}

\newcommand{\indi}{\nu_1\cdots\nu_{p+1}}

\newcommand{\indii}{\nu_1\cdots\nu_{p+2}}
\begin{document}
\begin{titlepage}
\begin{flushleft}
       \hfill                      {\tt hep-th/9907143}\\
       \hfill                       FIT HE 7-01/99 \\
\end{flushleft}
\vspace*{3mm}
\begin{center}
{\LARGE Yang-Mills theory\\ }
\vspace*{5mm}
{\LARGE from non-critical string\\ }
\vspace*{12mm}
{\large Kazuo Ghoroku\footnote{\tt gouroku@dontaku.fit.ac.jp}}\\
\vspace*{2mm}

\vspace*{2mm}

\vspace*{4mm}
{\it\large Fukuoka Institute of Technology, Wajiro, Higashi-ku}\\
{\it\large Fukuoka 811-0295, Japan\\}
\vspace*{10mm}
\end{center}

\begin{abstract}

The correspondence of the non-critical string theory 
and the Yang-Mills theory is examined according to the recent 
Polyakov's proposal, and two possible solutions
of the bulk equations are addressed
near the fixed points of the pure Yang-Mills theory: 
(i) One solution asymptotically approaches to the AdS space at the ultraviolet
limit where the conformally invariant field theory is realized. 
(ii) The second one approaches to the flat space 
in both the infrared and the ultraviolet limits. The area law
of the Wilson-loop and the asymptotic freedom with logarithmic behaviour
are seen in the respective limit.

\end{abstract}
\end{titlepage}

\section{Introduction}

~~Inspired by the conjecture \cite{M1,GKP1,W1} 
that the conformal symmetric super Yang-Mills theory is 
dual to the type IIB supergravity
in $AdS_5\times S^5$, several quantities 
of the Yang-Mills theory have been calculated in terms of the string theory
with D3-branes. There are two directions to extend
this duality to the non-supersymmetric gauge
theory. One way is to compactify one dimension to obtain a finite
temperature gauge theory \cite{W2}, 
and we get a non-supersymmetric theory. The 
running coupling constant 
is obtained by regarding the radius of the compactification as
the cut-off. Although
this formulation yields a successful result for
the area law of the Wilson-loop,
the difficulty appears in taking
the continuum limit where the temperature is taken as infinite.
Since the coupling constant in this formalism is proportional to
the temperature, then the tension-parameter of the QCD string exceeds the 
Planck mass. 

An alternative approach has been proposed by \cite{Poly1} 
in terms of the non-critical string theory which is constructed from
the supersymetric Liouville theory. In this theory, the space-time fermions
are removed by taking the GSO projection as in the type 0
string theory \cite{DH,SW}. As a result, two kinds of R-R fields and also
the tachyon field appear in the bulk space
other than the usual bosons of the type II theory. While 
in the world volume of the D-brane, the tachyon can be removed
by imposing the GSO projection on the
open string sector. Then
only the Yang-Mills field and some scalar fields exist in the
D-brane world volume, and the mass scale 
of the gauge theory is given by the Liouville field.

This formulation however suffers from a possible instability due to
the bulk tachyon when the dimension is higher than two.
This could be avoided by considering an appropriate potential 
\cite{CST,gho} which
leads to the condensation of the tachyon
or a curved space-time \cite{Bre}. An interesting idea of
the resolution for
this problem has recently be given by considering the couplings of
the R-R fields and the tachyon \cite{KT,KT3}.
Non-perturbative approaches
would however be necessary to obtain a satisfactory resolution of this problem
since we must know the correct
functional form of the action depending on the tachyon to solve the
problem of its vacuum.

While the idea of \cite{Poly1} was extended to type 
IIB ~\cite{NO,KS1,KS2,Gub,GPPZ,CM}
and type 0B \cite{KT,KT3,Min,Min2,KT4} models
by regarding the fifth coordinate of $AdS_5$ as the energy-scale
of the gauge theory. In type IIB model, asymptotic freedom 
of the gauge coupling has been shown
by adding the axion field to the effective supergravity action
\cite{KS2}, but it does not show the logarithmic
decreasing of the gauge-coupling. On the other hand,
the asymptotic freedom with the logarithmic behavior was seen
in the case of the type 0 model, but
it seems to be difficult to connect ultraviolet solution and the favorite 
infrared solution \cite{Min2}.

As indicated in \cite{Poly1}, the non-critical string theory also
has a solution corresponding to the conformal invariant gauge theory. 
Around this fixed point, some approaches to solving the renormalization
group equations have been given
in the scheme of the non-critical string theory \cite{FM,FM2,AFS,AG}, but
more considerations would be needed 
since the asymptotic freedom and other important properties of the gauge theory
are still obscure.

The purpose of this paper is to study furthermore the
running coupling constant and the Wilson loop of the gauge theory in 
terms of
the non-critical string theory, in which it is possible to study the pure 
Yang-Mills theory. Here we point out that there are two kinds of flows
of the renormalization group equations in the pure Yang-Mills theory.
They are specified by two types of asymptotic forms of the
bulk solutions, i.e. the 
AdS space and the flat space. The latter type solution shows
the asymptotic freedom with logarithmic decreasing for the coupling constant
in the ultraviolet region
and the area law of the Wilson-loop in the infrared limit. These behaviours
are shown by imposing some conditions on the tachyon potential and the
couplings of the tachyon to the R-R fields. But both kinds of
solutions are characterized by the different expectation value of the 
tachyon since the conditions imposed on the potential on this point are not
compatible. 

In section two we give the gravitational equations to be solved
and the conformal invariant interacting solution. 
The running behaviours from this fixed point are discussed
in the section three, and we find that the above
fixed point is the ultraviolet one smoothly connected to the solution given
in type II theory.
In section four, we show another solution with the asymptotic freedom 
near the ultraviolet fixed point. The gauge coupling decreases logarithmically
with the energy scale, and this solution would be connected to the one
given in the infrared limit. From this solution, we can see
the area law of the Wilson-loop. In the final section, the concluding remarks
are given.

\section{The gravitational equations and conformal fixed point}

~~The effective action of the non-critical string theory, whose boundary
represents the gauge theory, should include R-R $p+1$
potential $A_{p+1}$
\footnote{Although four kinds of the fundamental D-brane can be considered
\cite{KT}, we here consider only the electric type of R-R charge.
} 
other than the usual NS-NS fields. And we expect
that N $D_{p+1}$-branes are stacked on the boundary to make
the U(N) gauge theory there. 

Then we start from the following action,
\beqs
    S_D &=& {1\over 2\kappa^2}
      \int dx^D\sqrt{|g|}\Bigg\{e^{-2\Phi}\left(R-
      4(\nabla\Phi)^2+(\nabla T)^2+V(T)+c \right) \nonumber \\ 
    &{}&  \qquad\qquad~~~~~~ + \frac{1}{2(p+2)!}f(T)F_{p+2}^2\Bigg\},
    \label{action} 
\eeqs
where $c=-(10-D)/2\alpha'$, and $F_{p+2}=dA_{p+1}$ is the 
field strength of $A_{p+1}$. The total dimension
$D$ includes the Liouville direction, which is denoted by $r$.
The tachyon potential is represented by
$V(T)$, and $f(T)$ denotes the couplings between the tachyon and the R-R
field investigated in \cite{KT}. We consider their simple
forms in the next section. It should be noticed that there is a problem
of the stability of the tachyon
in extending the non-critical string theory to the higher dimension $(D>2)$.
In order to evade this problem, we assume the condensation of the 
tachyon and/or the resultant curved space which includes the asymptotically 
AdS space. In the case of AdS space, the tachyon can be stabilized
even if it has a small negative mass-squared \cite{Bre}. 
These points would be
cleared by using the exact forms of $V(T)$ and $f(T)$, but
we do not know them. So the problem of the stability 
of the tachyon is open here.

The equations of motion are written as
\beqs
     R_{\mu\nu} - 2 \nabla_\mu \nabla_\nu \Phi &=& -\nabla_\mu T\nabla_\nu T + 
             e^{2\Phi}f(T) T_{\mu\nu}^A \label{metric}\\
    4 \nabla_\mu \Phi \nabla^\mu \Phi 
       -2 \nabla^2 \Phi &=& \frac{D-2d-2}{4(p+2)!}e^{2\Phi}f(T)F_{p+2}^2
            + V_c(T) \label{dilaton}\\
    \nabla^2 T -2 \nabla_\mu \Phi \nabla^\mu T 
         &=& \frac{1}{2}V'_c(T)+\frac{1}{4(p+2)!}e^{2\Phi}f'(T)F_{p+2}^2 \,
            \label{tachyon}
\eeqs
\beq
     \partial_\mu (\sqrt{|g|}f(T)F^{\mu\indi}) = 0 \, , \label{form}
\eeq
where $V_c(T)= c+V(T)$ and
\beq  
 T_{\mu\nu}^A = -\frac{1}{2(p+1)!}
                \left(F_{\mu\indi}F_\nu^{\;\;\;\;\indi}
       - \frac{g_{\mu\nu}}{2(p+2)}  F_{\indii}F^{\indii}\right).
     \label{tensor3} 
\eeq

The advantage of the non-critical string scheme is to be able to 
avoid the extra dimensions of the bulk space. These freedom are 
corresponding to the adjoint scalars in the gauge theory, and they
are not necessary to see
the characteristic features of the Yang-Mills theory.
So the dimension $D$ is set as $D=p+2$ to consider the case of 
the pure Yang-Mills theory hereafter. But the extra
dimensions seem to be
essential to obtain an asymptotic free solution given in the
type 0 model. We comment on this point
in the proceeding sections.

We solve the above equations according to the following ansatz;
\beq
 ds^2= e^{2A(r)}\eta_{\mu\nu}dx^{\mu}dx^{\nu}
           +e^{2B(r)}dr^2 \, \label{metrica}
\eeq
\beq
    \Phi\equiv\Phi(r), \quad \quad T\equiv T(r) \quad \hbox{and} \quad
     A_{01\cdots p} = -e^{c(r)}\ ,
\eeq
where $x^{\mu}, \mu=0\sim p$, denote the space-time coordinates,
and $r$ represents the Liouville direction. Here
$r$ plays the role of the energy scale for the $p+1$-dimensional gauge theory
\cite{Poly1}.
The equation (\ref{form}) is solved as
\beq
 \partial_{r}e^{c(r)}={N\over f(T)} e^{dA+B} \, \label{RRfi}
\eeq
where $d=p+1$ and N denotes the number of the p-brane.
Then the remaining equations (\ref{metric}) and (\ref{dilaton}) are 
rewritten as,
\beq
    -\dot{A}\dot{B}+\ddot{A}+d\dot{A}^{2}-2\dot{A}\dot{\Phi}=
       e^{2B+2\Phi}{N^2\over 4f(T)}, \label{metric1}\\
\eeq
\beq
       d(\ddot{A}+\dot{A}^2-\dot{A}\dot{B})
          -2(\ddot{\Phi}-\dot{B}\dot{\Phi})=
       e^{2B+2\Phi}{N^2\over 4f(T)}-\dot{T}^2,  \label{metric2}\\
\eeq    
\beq
         2(\ddot{\Phi}+\dot{\Phi}(d\dot{A}-\dot{B}))-4\dot{\Phi}^2
      = e^{2B+2\Phi}{(d+1)N^2\over 4f(T)}+e^{2B}V_c(T) \,
                 \label{dilaton2}\\
\eeq
\beq
    \ddot{T}+(d\dot{A}-\dot{B})\dot{T}-2\dot{\Phi}\dot{T}=
        -{f'(T)\over 2f^2(T)}N^2e^{2B+2\Phi}+{1\over 2}e^{2B}V'_c(T) \,
         \label{tachyon2}
\eeq
where $\dot{}$ denotes the derivative with respect to $r$.

Before considering the problem of the renormalization group flow
of the gauge theory, we give the anti-de 
Sitter solution as a conformal invariant fixed point since it is the 
starting point of our study. It can be found by solving the above equations
by assuming that $\Phi$ and $T$ are constant, $i.e.$ independent on $r$,
and taking the following ansatz,
\beq
  e^A=({r\over r_0})^{\gamma}\
       , \ \quad e^B=({r\over r_0})^{b}\ , \label{ansatz}
\eeq
where $r_0$ denotes a scale parameter which measures the scalar curvature
(see (\ref{curva})). After a small calculation, we get
\beq
 b=-1\,  , \label{ads}
\eeq
and
\beq
   V_c(T_0)+{d+1\over 4f(T_0)}\lambda_0^2=0 \ , 
      \quad  \gamma^2= {r_0^2\over 4df(T_0)}\lambda_0^2 \, , \label{dil}
\eeq
\beq
   V_c^{'}(T_0)-{f'(T_0)\over f^2(T_0)}\lambda_0^2=0 \ , \label{tach}
\eeq
where $\lambda_0=Ne^{\Phi_0}$ is the t'Hooft coupling constant. 
Here we notice that the scale parameter $r_0$ is related
to the scalar curvature $R$ as
\beq
  R\equiv g^{\mu\nu}R_{\mu\nu}=(1+d)d{\gamma^2\over r_0^2} \ , \label{curva}
\eeq
which is derived from Eqs. (\ref{ansatz}) and (\ref{ads}).

From Eqs. (\ref{ads}) $\sim$ (\ref{tach}), we can
determine $T_0$, $\lambda_0$ and $\gamma$ if the explicit forms of $V_c(T)$
and $f(T)$ are given. As an example, 
we solve them by using the following simple forms 
\beq
  V_c(T)=-{9-d\over 2}-{T^2\over 2}\, , \qquad f(T)=1+T+{T^2\over 2}\, , 
        \label{tdep}
\eeq
where we take as $\alpha'=1$, and the above forms are obtained from the 
expansion of the effective action near $T=0$ \cite{KT}.
Then the solutions are given as 
\beq
  \lambda_0=1.17, \quad T_0=-0.814, \quad \gamma =0.753 . \label{solo}
\eeq
We can see the deviation of the value of $\gamma$ from one, which
is obtained in the type IIB model.

Next, we briefly comment on the formulation based on the thermalization, where
the Hawking temperature $T_H$ is introduced
as a cut-off parameter of the gauge theory according
to \cite{W2}.
The equations are solved by the following ansatz with
the thermal deformation $h(r)$,
\beq
 ds^2= -e^{2A_0(r)}dx^{0}dx^{0}+e^{2A(r)}\delta_{ij}dx^{i}dx^{j}
           +e^{2B(r)}dr^2 \, \label{metricb}
\eeq
where $i, j=1\sim 3$ and
\beq
  e^{A_0}=h(r)({r\over r_0})^{\gamma_0}\ ,\quad e^A=({r\over r_0})^{\gamma}\
       , \ \quad e^B=h^{-1}(r)({r\over r_0})^{b}\ . \label{metrict}
\eeq
After a calculation, we find the solution eqs.(\ref{ads}), (\ref{dil}), 
(\ref{tach}) and 
\beq
   h(r)=1-({r_1\over r})^{d\gamma} \ , \label{therma}
\eeq
where $r_1$ denotes a new scale parameter. And $T_H$ can be derived 
from the completeness of the metric as follows,
\beq
  T_H={d\gamma\over 4\pi r_0^{\gamma +1}}r_1^{\gamma}\ \ .\label{tempe}
\eeq
In this construction, we can see the well-known 
thermodynamical properties of the 
Yang-Mills gas with the temperature $T_H$, but it is difficult to 
connect this solution to the
asymptotic free running coupling constant,
$\lambda_p$, of $p$-dimensional gauge theory.
It is given as,
\beq
 \lambda_p=\lambda_0 T_H \, .
\eeq
Since $\lambda_0$ is a cnstant and $T_H\to \infty$ in the ultra-violet limit,
then $\lambda_p$ grows with $T_H$. So it seems necessary
to consider other formulation of the conformal breaking. 
As discussed below, one direction is to solve the equations by
considering the $r$ dependence of the fields $\Phi$ and $T$, which are
regarded as the running coupling constants.

\section{Asymptotic CFT solution}

First, we give the solution whose ultraviolet limit is given by 
the AdS solution given in the
previous section with constant $\Phi$ and $T$. Here the $r$-dependence of 
$\Phi$ and $T$ are turned on, and the equations are solved by
adopting the ansatz (\ref{metrica}) for the metric
and the following form of the solution,
\beq
  A(r)=\gamma\ln(r/r_0)+a(r), \qquad B(r)=-\ln(r/r_0)+b(r), \label{an1}
\eeq
\beq
  \Phi(r)=\Phi_0+\phi(r), \qquad T(r)=T_0+t(r), \label{an2}
\eeq
where $\gamma, \Phi_0$ and $T_0$ are determined by the equations (\ref{ads}), 
(\ref{dil}) and (\ref{tach}). The deviations, 
$ \{a(r), b(r), \phi(r), t(r)\}$, from the AdS limit are 
obtained by solving Eqs. (\ref{metric1}) $\sim$ (\ref{tachyon2}). 
Since it is difficult to find an exact solution, we solve the
equations by expanding the deviations in power 
series of $(r_0/r)^{\alpha}$ as
\beq
  \chi^i(r)=\sum_n \chi^i_n({r_0\over r})^{n\alpha} \, , \label{an3}
\eeq
where the deviations defined in (\ref{an2}) are denoted by a
vector $\chi^i(r)\equiv$ $(a(r),$ $ b(r), \phi(r), t(r))$.
For example, the first component represents 
$a(r)=\sum_n$ $a_n({r_0\over r})^{n\alpha}$.
The coefficients $\chi^i_n$ ( $a_n$ $etc.$ )
and the index $\alpha$ are determined by solving the equations.
Here we notice two points: 
(i) The reason why $\alpha =1$ is not considered is that the index $\alpha$
corresponds to the lowest order coefficient of the 
$\beta$-function of the gauge coupling as
seen in Eq. (\ref{beta}). In general this coefficient is determined
by the dynamical property of the system.
(ii) The expansion form (\ref{an3}) can be used for both the ultraviolet and
infrared regions for the positive and negative value of $\alpha$ respectively.

The solution of the lowest order equations is equivalent to Eqs. (\ref{dil})
and (\ref{tach}). From the first order equations of $O(r^{-\alpha})$, 
the following condition is obtained if there is a non-trivial solution,
\[
 [\alpha(\alpha-d\gamma)]^2+\alpha(\alpha-d\gamma)
 ({1\over 2}V(T_0)r_0^2-f_2) \nonumber
\]
\beq
~~~~~~~~~~~ -{1\over 2}V(T_0)r_0^2 f_2+V'(T_0)r_0^2 f_1=0~, \label{det}
\eeq
where 
\beq
f_1={1\over 4}V(T_0)r_0^2(V'(T_0)/V(T_0)+f'(T_0)/f(T_0))~ .
\eeq
\beq
f_2={1\over 2}V(T_0)r_0^2[V''(T_0)/V'(T_0)-f''(T_0)/f'(T_0)+2f'(T_0)/f(T_0)].
\eeq
Here the Eq. (\ref{det}) comes from the Eqs. (\ref{dilaton2})
and (\ref{tachyon2}) as a condition that $\phi_1$ and $t_1$ are not zero.
If we take $\phi_1=t_1=0$, then $a_1=b_1=0$ are obtained 
from Eqs. (\ref{metric1}) and
(\ref{metric2}). And this is the trivial solution. 
Since $T_0$ and $\gamma$ are
determined from Eqs. (\ref{dil}) and (\ref{tach}) if $V(T)$ and $f(T)$ are
given, then the value of $\alpha$ is given by solving Eq. (\ref{det})
for non-trivial solutions. We notice here 
that Eq. (\ref{det}) generally has four solutions for $\alpha$,
but there is no principle to choose one of them at present.
Here, we adopt the solution which is smoothly
continued to the one of IIB model given
at $D=10$ and $d=4$. In this case, the 
right hand side (r.h.s.) of Eq. (\ref{dilaton}) is zero, then
the corresponding solution
is obtained from Eq. (\ref{dilaton2}) by equating its r.h.s. to be 
zero. From the condition that $\phi_1\neq 0$, we obtain
\beq
   \alpha=d\gamma \, .  \label{alpha}
\eeq
In type IIB model \cite{KS1}, $\gamma=1$ and $\alpha=4$ has been obtained.
In the non-ctitical case, we take the above form of Eq. (\ref{alpha}) 
as the solution
for $\alpha$, but the value of $\gamma$ is different from the one of the
type IIB case. The value of $\gamma$ is given by the solution of the zeroth
order equations, which depends on the form of $V_c(T)$ and $f(T)$. Their
simple form are given in the previous section, but we must here 
impose a condition to them such that we can find
the solution (\ref{alpha}). Substituting this solution
into (\ref{det}), the next condition is obtained
\beq
  ({V_c'(T_0)\over V_c(T_0)})({V_c'(T_0)\over V_c(T_0)}+{f'(T_0)\over f(T_0)})=
    {V_c''(T_0)\over V_c'(T_0)}
      -{f''(T_0)\over f'(T_0)}+2{f'(T_0)\over f(T_0)} .
            \label{cond}
\eeq

It is easily seen that the simple form of $V_c(T)$ and $f(T)$ given in the 
previous section (Eq. (\ref{tdep}))
give two real solutions $\alpha=3.43$ and $-0.416$ with $T_0=-0.814$, but
$d\gamma=3.01$ so these solutions
do not satisfy the above condition. In order to get the solution
$\alpha=d\gamma$, we must find other form of (\ref{tdep}) such that they 
satisfy Eq. (\ref{cond}). It is straightforward to determine the 
coefficients of the expansions by solving the equations if the explicit
forms of $V_c(T)$ and $f(T)$ are given. But
to find such functions is out of our present work,
so it would be addressed in a separate paper.

We comment on the meaning of $\alpha$ from the viewpoint
of the Yang-Mills theory. 
The solutions of $\alpha=d\gamma >0$ can be regarded
as the non-critical string version of the one given
in the type IIB model \cite{KS1}. As in \cite{KS1},
we can see the renormalization group equation of
the gauge coupling constant, $\lambda=N e^{\Phi}=Ng_{YM}^2$. It is
expanded as
\beq
  \lambda = \lambda_0 \left\{1+\phi_1({r_o\over r})^{\alpha}+\cdots\right\},
\eeq
then we obtain from the definition of the $\beta$-function, 
$\beta(\lambda)\equiv rd\lambda/dr$, the following result:
\beq
 \beta'(\lambda_0)= -\alpha  . \label{beta}
\eeq
From this, we can see that the AdS limit is the ultraviolet 
(infrared) fixed point for $\alpha >0$ ( $\alpha <0$ ).
Although the result is dependent on the form of $V_c(T)$ and $f(T)$,
it is expected that the AdS is the ultraviolet fixed point if it is
smoothly connected to the type IIB model.

Nextly we consider the quark-antiquark potential ($U_{Q\bar{Q}}$) 
with a distance $L$ between them. The potential can be obtained
by estimating the Wilson-loop according to \cite{M2}.
The minimized Nambu-Goto action, $S_m$, and 
the distance $L$ can be obtained by using
the approximate solution with the first order corrections as
\beq
  S_m={\tau\over 2\pi}E^{1/2}{r_0\over 4\gamma}
      [B(-1/4,1/2)+(a_1+b_1)E^{-\alpha/2\gamma}B(-1/4+\alpha/4\gamma,1/2)] ,
\eeq
\beq
  L= E^{-1/2}{r_0\over 4\gamma}
      [B(3/4,1/2)+(b_1-3a_1)E^{-\alpha/2\gamma}B(3/4+\alpha/4\gamma,1/2)] .
\eeq
Here $B(a,b)$ denotes the beta-function and $E$ is an
integral constant. And $\tau$ is the length in the time-direction of the
Wilson loop. The above formula are obtained from the approximate 
solution of the order
$O((r_0/r)^{\alpha})$, so we write the resultant potential derived from
them as the following approximate form,
\beq
  U_{Q\bar{Q}}={1\over 2\pi}c_3{r_0\over L}
         (1+c_2({L\over c_1r_0})^{\alpha/\gamma}
            +\cdots ),
\eeq
where 
\beq
   c_1= {1\over 2\gamma}B({3\over 4},{1\over 2}), 
       \quad c_3={c_1\over 4\gamma} B(-{1\over 4},{1\over 2}) ,
\eeq
\beq
   c2=(b_1-3a_1){B({3\over 4}+{\alpha\over 4\gamma},{1\over 2}) \over 
         B({3\over 4},{1\over 2})} +
         (a_1+b_1){B(-{1\over 4}+{\alpha \over 4\gamma},{1\over 2})
           \over B(-{1\over 4},{1\over 2})}.
\eeq
For $c_3>0$, Coulomb attraction can be seen near the fixed point as in type
IIB model.
The next order correction in $ U_{Q\bar{Q}}$ would deviate from
the result of \cite{KS1} since $\alpha /\gamma \neq 4$ is expected.

The solutions obtained here can be regarded as fluctuations around the AdS
space near its limit, 
and they are the renormalizable one. Then the solutions represent
the renormalization group flow of the theory, which is existing on the
boundary of AdS, with non-zero expectation value of some operator in the
theory \cite{BKL,BKLT,GPPZ}. This point was discussed also for type II theory
in the first paper of \cite{NO}. The CFT on the boundary is the $N=4$
super Yang-Mills theory in the case of type IIB, but it would not be a
sypersymmetric theory in our case since our model is based on the type
0 theory in non-critical dimension. In any case,
the solutions considered in this section would represent the running
behaviour of the parameters in a different phase from that we are
searching for. In the next section, we discuss the 
more desirable solution which shows
the asymptotic freedom at ultraviolet region.

\section{Asymptotic free solution}

Here we investigate the solutions which leads to the asymptotic freedom of the
gauge coupling constant, i.e. it
decreases logarithmically with the energy scale. The equations in the 
ultraviolet region (large $r$) are
solved in terms of the following expansions for $\chi^i$ 
according to \cite{Min2,KT3},

\beq
  a(r)=\bar{a}_0\ln y+\bar{a}_1{1\over y}(\ln y+\bar{a}_{10})
         +\bar{a}_2{1\over y^2}(\ln^2y+\bar{a}_{21}\ln y
       +\bar{a}_{20})+\cdots  , \label{asyma}
\eeq
\beq
  b(r)=\bar{b}_0\ln y+\bar{b}_1{1\over y}(\ln y+\bar{b}_{10})
         +\bar{b}_2{1\over y^2}(\ln^2y+\bar{b}_{21}\ln y
       +\bar{b}_{20})+\cdots  , \label{asymb}
\eeq
\beq
  \phi(r)=\bar{\phi}_0\ln y+\bar{\phi}_1{1\over y}\ln y
       +\bar{\phi}_{2}{1\over y^2}\ln^2y+\cdots , \label{asymc}
\eeq
\beq
  t(r)= \bar{t}_1{1\over y}+
         \bar{t}_2{1\over y^2}(\ln y
       +\bar{t}_{20})+\cdots  , \label{asymd}
\eeq
where $y=\ln(r/r_0)$, and
we assume $\bar{\phi}_0<0$, which is required from the asymptotic 
freedom. 

While the equations (\ref{metric1}) $\sim$ (\ref{tachyon2}) 
are rewritten by using $\chi^i$ as follows:
\beq
      \ddot{a}+\dot{a}(d\dot{a}-\dot{b}-2\dot{\phi})+
       \gamma(2d\dot{a}-\dot{b}-2\dot{\phi}) 
  =-d\gamma^2+{r_0^2\lambda_0^2\over 4f(T)}
       e^{2(b+\phi)}, \label{metric11}
\eeq
\beq
       d[\ddot{a}+\gamma(2\dot{a}-\dot{b})
                +\dot{a}(\dot{a}-\dot{b})]
           -2(\ddot{\phi}-\dot{b}\dot{\phi})+\dot{t}^2
 =-d\gamma^2+{r_0^2\lambda_0^2\over 4f(T)}
       e^{2(b+\phi)}, 
                \label{metric21}
\eeq    
\beq
    \ddot{\phi}+\dot{\phi}(d\dot{a}-\dot{b}-2\dot{\phi})+d\gamma\dot{\phi}
     = (d+1){r_0^2\lambda_0^2\over 8f(T)}
       e^{2(b+\phi)} +{r_0^2V_c(T)\over 2}e^{2b} \, ,
                 \label{dilaton21}
\eeq
\beq
    \ddot{t}+\dot{t}(d\dot{a}-\dot{b}-2\dot{\phi})+d\gamma\dot{t}
      = -{r_0^2\lambda_0^2f'(T)\over 2f^2(T)}
       e^{2(b+\phi)} +{r_0^2V_c'(T)\over 2}e^{2b} \, ,
         \label{tachyon21}
\eeq
where the dot denotes the derivative with respect to $y$.

For this formulation, we firstly find
\beq
   \gamma=0 \, . \label{lack}
\eeq
This is seen as follows. From Eqs. (\ref{metric11}) and (\ref{metric21}), 
we must require the relation, $\bar{b}_0=-\bar{\phi}_0>0$, to obtain the result
$\gamma\neq 0$. If we use this relation in Eqs. (\ref{dilaton21}) and 
(\ref{tachyon21}), then we obtain $V_c(T_0)=V_c'(T_0)=0$. Then $\lambda_0=0$
is derived from the same equations, but this result is not 
compatible with $\gamma\neq 0$ from Eqs. (\ref{metric11}) and (\ref{metric21}).
Then we are led to the result (\ref{lack}). 
This result implies that the bulk space is asymptotically not
the AdS but the flat space as seen from (\ref{curva}). 
This seems to be reasonable
since the flux of the R-R field, which was the essencial factor to derive
the AdS space, is suppressed in the equations to be solved
by the factor $e^{2\Phi}$ near the ultraviolet
limit due to the asymptotic freedom. And the flux term contributes to 
the higher orders of equations to reform the flat space to a curved one.

We notice however that there are two ways to get an asymptotic free 
solution which is compatible with the asymptotic AdS bulk-space. One 
way is to consider the extra dimension with the following metric,
\beq
 ds^2= e^{2A(r)}\eta_{\mu\nu}dx^{\mu}dx^{\nu}
           +e^{2B(r)}dr^2+e^{K(r)}\hat{g}_{ab}dx^adx^b \, , \label{metric3}
\eeq
where $\hat{g}_{ab}$ denotes (D-d-1)-dimensional sphere. In this case,
the solution for the RR-field (\ref{RRfi}) is altered and
the factor $e^{2(b+\phi)}$ of the R-R flux term in the equations is changed
to $e^{2(b+\phi-[D-d-1]K)}$. Then the R-R term in the lowest order of
equations can be retained even in the asymptotic-free case
since the energy dependence of $\phi$ can be
cancelled by $b$ and $K$. This was explicitly seen in the Einstein frame
of the type-0 model \cite{Min2}.
As a result, one can get an asymptotic AdS solution with the asymptotic
freedom. However this is restricted to the case of $D=10$ for $p=3$
due to the consistency of the equations, which demands the condition,
$D-2p-4=0$. Then this type of solutions are excluded
for the pure Yang-Mills theory, in which
$D=5$ for $p=3$, and also for $D<10$ with the extra spaces $x^a$.

The second way is to consider a special form of $f(T)$, which could
cancel the logatithmic energy dependence of $e^{2(b+\phi)}$. However
this idea seems to be unreasonable since the tachyon should largely
deviate from $T_0$  at large $y$ due to the factor $\ln y$ which is necessary
to cancel the same factor in $b+\phi$. 
As a result, one expects that the bulk system would become unstable
there. 

Then it is the most probable case to consider
the suppression of the RR term in the ultraviolet limit.
This implies that the asymptotic bulk space in the non-critical string theory
is not the AdS but the flat space which is dual to the asymptotic-free 
Yang-Mills theory. Then we firstly solve the equations by neglecting 
the R-R term in the lowest order. While
the order of this R-R term is determined by solving the equations of
lower order series where this term is still neglected.

On the other hand, there is no reason to suppress $V_c(T)$ in the lowest
order of equations, and the situation depends on the value of $\bar{b}_0$.
There are two possibilities, $i.e.$ (i) $\bar{b}_0=-1$ and (ii)
$\bar{b}_0<-1$. In the first case, $V_c(T)$ should be retained.
And $V_c(T)$ is also neglected in the second case, but it is seen 
in this case that we can not determine
the value of $\bar{b}_0$ and $\bar{\phi}_0$ by the lower order of equations
at least up to $O(y^{-3})$. 
Then we cosider the former case, $\bar{b}_0=-1$, hereafter. 

In this case,
the potential term $V_c(T)$ is included in the lowest order equations
of (\ref{dilaton2}) and (\ref{tachyon2}). There are two possible solutions
$\bar{\phi}_0=-1/2$ and $\bar{\phi}_0=-1$. Here we adopt the latter solution
since it leads to the correct index of the logarithmic decreasing of the
gauge coupling, $i.e.$ $g^2_{YM}\propto 1/y$. 
This means that the R-R term appears 
in the equations of order $O(1/y^{4})$ for the first time.
Then the followings are obtained by solving the 
equations up to the order of $O(1/y^{4})$:
\beq
  a(r)= {\bar{\lambda}_0^2\over 4}{1\over y^2}\ln y+\cdots , \qquad
  b(r)= -\ln y-{\bar{\lambda}_0^2\over 2}{1\over y^2}(\ln^2 y+\ln y)+\cdots ,
 \label{asyma1}
\eeq
\beq
  \phi(r)= -\ln y - {\bar{\lambda}_0^2\over 4}{1\over y^2}\ln^2 y+\cdots,
    \qquad 
  t(r)= O({1\over y^3}) , \label{asymd1}
\eeq
and
\beq
 r_0^2 V_c(T_0)= -4, \qquad 
  V_c'(T_0)= 0 ,   \label{potco2}
\eeq
where $\bar{\lambda}_0^2=\lambda_0^2/f(T_0)$.
The coefficients depending on $\bar{\lambda}_0^2$ are coming
from equations of order $O(1/y^{4})$.

Other coefficients are given by solving the equations of higher orders, but
we do not do it here. 
For the solution up to this order, we can make the following remarks:
(i) The asymptotic bulk geometry is the flat space. This is seen by changing 
the Liouville coordinate $r$ to $\rho\equiv \ln\ln r$, and also from Eq.
(\ref{curva}) for $\gamma=0$.
(ii) The Yang-Mills coupling constant, which is defined by
 $g_{\rm YM}^2=e^{\Phi}$, decreases like $g_{\rm YM}^2\propto (\ln r)^{-1}$
at large $r$. And this index $-1$ is the expected value from the perturbative
calculation in the Yang-Mills theory. But there is no term like $y^{-1}\ln y$
in $\phi(r)$, 
so the two-loop order correction for $g_{\rm YM}^2$ is lacking in this result.
(iii) The value of the potential at $T_0$ is restricted
through the conditions in (\ref{potco2}). We can see that 
the latter condition $V_c'(T_0)=0$ is not compatible with the condition
required in the section 3 to obtain the AdS fixed point.
Then the asymptotic free
solution would represent the different phase of the gauge theory
from the one given in the previous section.

Related to the second remark, we notice that
the expected value of index is not obtained in the type 0B model.
But this problem is resolved in 0B model
by considering the effective coupling-constant obtained by evaluating the 
Wilson-Loop \cite{Min2} near the ultraviolet region.

In our model, the leading part of the $Q\bar{Q}$ potential can be 
estimated by using the lowest order of solution:
\beq
  A(r)=0, \qquad B(r)=-\ln(r/r_0)+b_0\ln\ln(r/r_0). \label{sol2}
\eeq
According to the usual analysis, we obtain
\beq
 U_{Q\bar{Q}}=S_0\sqrt{1+({L\over L_0})^2}~, \label{qpot2}
\eeq
where 
\beq
 S_0={\tau\over 2\pi}L_0~, 
  \qquad L_0=r_0\int_{\epsilon}^{\infty}dy y^{b_0}~.
\eeq
Here we introduced an infrared cutoff $\epsilon$ since the approximate
formula (\ref{sol2}) is useful for large $r$ (large $y$). Since we are
considering the ultraviolet region, then (\ref{qpot2}) would be useful
at small $L$ and
it should be expanded as follows,
\beq
 U_{Q\bar{Q}}\sim S_0[1+{1\over 2}({L\over L_0})^2]~. \label{qpot21}
\eeq
This represents the harmonic type of potential, and
the Coulomb potential does not appear. The reason why we can not see
the $1/L$ term is that our solution is not the asymptotic AdS one, which
leads to a conformal field theory as a fixed point and the Coulomb potential.
While, a Coulomb potential with a logarithmic correction was found in the 
type 0B model. This is because of that
the bulk solution is asymptotically AdS in this case. But this
kind of behaviour can not be seen in the pure Yang-Mills theory
because the bulk space is asymptotically
flat. So we can not discuss the problem of the index of the logarithmic
behaviour of the coupling constant as in the type 0B model.

Although Eq. (\ref{qpot21}) is valid at small $L$, this potential implies
the quark confinement. In fact, the Eq.
(\ref{qpot2}) shows the linear potential at large $L$. So we examine whether
the similar potential can be obtained also in the infrared region.

We can proceed the analysis with the same formalism in the infrared region.
First, the variable $y$ is changed to $y=-\ln r$ and 
the region of large $y$ ( small $r$) is considered. 
Then, Eqs. (\ref{metric11}) $\sim$ (\ref{tachyon21}) used in the ultraviolet 
region are also useful in the infrared region 
if we change the sign of the single derivative terms like
$\dot{a}\to -\dot{a}$. Then we can use the same expansion form of $\chi^i$,
(\ref{asyma}) $\sim$ (\ref{asymd}), for the infrared region. 

In solving the equations, we firstly consider the R-R term as in the 
ultraviolet limit. This term
is proportional to $e^{2(b+\phi)}$, so we expect that
it increases in the infrared region since $e^{\Phi}$ becomes large. 
Here we search for a solution which asymptotically approaches to 
the flat space also in the infrared limit, $i.e.$ $\gamma=0$. Because 
this solution could lead to a favourable $Q\bar{Q}$ potential.
In order to obtain this type
of solution, we suppose that $e^{\Phi}\to e^{\Phi_0}$ 
in the infrared limit and $b_0<-1$.
Namely, the R-R term is suppressed in the lowest order equations also
in the infrared limit, and the gauge coupling approaches a large but
finite-constant value.
Due to these restrictions, we are led to the solution
\beq
      a_0=0 ~, ~~\phi_0=0,
\eeq
by solving the lowest order equations of $O(1/y^2)$. 

From the next order of equations, 
we obtain $b_0=-2$. So the R-R term and the potential of $T$
appears in the equations of order
$O(1/y^4)$ for the first time. Up to this order,
we obtain
\beq
  a(r)=O({1\over y}), \qquad
  b(r)=-2\ln y+O({\ln y\over y}), \label{iasymb} 
\eeq
\beq
  \phi(r)=O({\ln y\over y^2}), \qquad
  t(r)= -2{f'(T_0)\over f(T_0)}{1\over y^2}+O(1/y^3) , \label{iasymd}
\eeq
and 
\beq
   \lambda_0^2=8f(T_0) . \label{coupling}
\eeq
The coefficients of the above terms written as $O(..)$ can not be 
obtained by the parameters given in
the equations up to this order, but they are taken to be non-zero values.
The last result (\ref{coupling}) is obtained by using Eq. (\ref{potco2})
obtained in the urtraviolet region by assuming that the two solutions
are connected smoothly. This implies that the coupling constant at the 
infrared fixed point is given by the coupling of the tachyon to the
R-R fields.

For this solution, we obtain the same $Q\bar{Q}$ potential with the one
obtained in the ultraviolet region, Eq. (\ref{qpot2}). It should be used for
large $L$ in this case, and we can see the linear potential,
\beq
 U_{Q\bar{Q}}\sim {\tau\over 2\pi}L~. \label{qpot22}
\eeq
From this, the QCD string tention is obtained as $1/2\pi$ in the unit of
$\alpha'=1$. This is
the expected behavior for the Yang-Mills theory in the infrared region.
It is important to find a solution which connects the solutions
given in the two limits, but it is open here.

\section{Conclusions}

 By extending the idea of the AdS/CFT correspondence, we have examined
the equations of the effective action of non-critical string theory
as the renormalization group equations of the Yang-Mills theory.
The analysis is restricted to the case of $D=5$, where
$D$ is the bulk dimension, to see the properties of the pure Yang-Mills
theory. In this model, the unknown functions of the tachyon are
included, so we need their explicit form if we want the quantitative
and definite
results. However, we could get several qualitative features of 
the Yan-Mills theory from the asymptotic solutions of 
the non-critical string. Two kinds of ultraviolet fixed points have been found;
(i) One is the asymptotic free fixed point with the logarithmic decreasing
of the gauge coupling constant with respect to the enrgy scale. 
The index of the lowest term of the gauge coupling is consistent with
the Yang-Mills theory, but the two loop correction term is lacking.
Applying this scheme to the small energy scale limit,
the solution near the infrared fixed point is also found.
(ii)
Another is the one of a conformal field theory limit
corresponding to the $AdS_5$ bulk space, which is related
to the solution of type IIB model if we impose 
an appropriate condition on the 
potential ($V_c(T)$) and the coupling to the R-R field ($f(T)$) of the tachyon
field ($T$) at $T=T_0$. 

The second solution approaches to the AdS near the fixed point, and 
Coulombnic $Q\bar{Q}$ potential is seen from 
the estimation of the Wilson-loop
by using the asymptotic solution. This solution can be considered as the
variation of the one given in the type IIB model. 
On the other hand, the asymptotic form of
the first type solution is the flat space, and this leads to the harmonic 
osccilator type $Q\bar{Q}$ potential in the ultraviolet limit. And the 
linear potential is obtained in the infrared region, where the gauge 
coupling approaches to a constant fixed point. Then it would be probable
that there is an exact solution which smoothly connect these two limit
solutions. It is an open problem to find such solution here.

The two urtlaviolet fixed points of type (i) and (ii) are realized at
different value of $T_0$, which is independent on the energy scale, so
these points could not be connected by one solution of our equations
presented here. New elements or new dynamical speculations would be
needed to connect them. This would be an interesting problem of the
pure Yang-Mills theory.

\vspace{.5cm}

\noindent {\bf Acknowledgement}

\vspace{.3cm}
  The author thanks to Profs. K. Inoue and G. Ferretti for their
useful comments.

\end{document}